\documentclass{article}

\usepackage[a4paper,hscale=0.77,vscale=0.82]{geometry}
\usepackage{amsmath,amsthm,amssymb}
\usepackage{array}
\usepackage{booktabs}
\usepackage{algorithm}
\usepackage{algorithmicx}
\usepackage[noend]{algpseudocode}

\newcommand{\ZZ}{\mathbb{Z}}
\newtheorem{theorem}{Theorem}
\newtheorem{lemma}{Lemma}

\begin{document}

\title{Secure data hiding for digital contact tracing}
\author{Craig Gotsman \and Kai Hormann}
\date{}
\maketitle


\begin{abstract}
Digital contact tracing is an effective tool in controlling the spread of infectious diseases such as COVID-19. It involves digital monitoring and recording of physical proximity between people over time with a central and trusted authority, so that when one user reports infection, it is possible to identify all other users who have been in close proximity to that person during a relevant time period in the past and alert them. One way to achieve this involves recording on the server the locations, e.g.\ by reading and reporting the GPS coordinates of a smartphone, of all users over time. Despite its simplicity, privacy concerns have prevented widespread adoption of this method. Technology that would enable the ``hiding'' of data could go a long way towards alleviating privacy concerns and enable digital contact tracing at a very large scale. In this article we describe a general method to hide data. By hiding, we mean that instead of disclosing a data value $x$, we would disclose an ``encoded'' version of $x$, namely $E(x)$, where $E(x)$ is easy to compute but very difficult, from a computational point of view, to invert. We propose a general construction of such a function $E$ and show that it guarantees perfect recall, namely, all individuals who have potentially been exposed to infection are alerted, at the price of an infinitesimal number of false alarms, namely, only a negligible number of individuals who have not actually been exposed will be wrongly informed that they have. Our encoding method does not require the use of public or private encryption keys, and its security relies on the sheer size of the relevant spatio-temporal data domain.
\end{abstract}


\section{Introduction}\label{sec:intro}

Contact tracing has proven to be an effective tool in controlling the spread of infectious diseases such as COVID-19. It involves investigating the movement and human contacts that an infected person has had in the days prior to the infection being discovered, and notifying and isolating these persons in the hope of stopping the spread. Obviously the process repeats if one of these persons has already been infected. While effective when done properly, the manual contact tracing process is time-consuming, tedious and error-prone, as not all contacts may be discovered and subsequently traced. It is estimated that a workforce of 100,000 ``contact tracers'' are required in the USA alone to cover the COVID-19 outbreak, yet only slightly more than 53,000 were active in October 2020, seven months after the pandemic began in the USA~\cite{NPR:2020:CCT}. It is estimated that once the infection rate hits 10 new cases per day per 100,000 people, manual contact tracing will become infeasible. The only hope for contact tracing at that point is to automate it by \emph{digital} monitoring and recording of physical proximity between people over time, so that when one user reports infection, it is possible to identify all other users who have been in close proximity to that person during a relevant time period in the past and alert them. These users would be required to monitor their symptoms and isolate, allowing early treatment and preventing further spread. Digital contact tracing (sometimes called \emph{automatic} contact tracing) was pioneered and deployed successfully in countries such as China, South Korea, Singapore, Israel, Australia, and Germany, and since the early days of the pandemic, many apps have been developed worldwide implementing digital contact tracing. The interested reader is referred to the survey of Ahmed et al.~\cite{Ahmed:2020:ASO} for a comprehensive description of many of them.

There are two main approaches to digital contact tracing. The first is based on the \emph{relative} distance between users. Using the Bluetooth sensor on a smartphone, it is possible to detect signals from other users with Bluetooth emitters who are physically close by (i.e.\ within a certain range) and record the proximity, either locally on the user's device, or at a central authority/server. This method, developed by Apple and Google in their \emph{Exposure Notification} (EN) framework~\cite{Apple:2020:PCT}, has the advantage that absolute locations of users are never disclosed, ensuring some degree of privacy. The disadvantage is the reliability of the Bluetooth sensors and their ability to work well under all relevant conditions (e.g.\ occlusion) and at all relevant ranges and some security concerns about the popular decentralized approach to storing this type of data on user devices \cite{vanBass:2021}. In retrospect, although digital contact tracing based on this technology held much promise, in practice it was plagued with operational issues and used by far fewer people than what would be required to make an impact \cite{NYT:COVID}.

The second approach to digital contact tracing involves recording on a central server the \emph{absolute} locations, e.g.\ by reading and reporting the GPS coordinates of a smartphone, of all users over time. This obviously provides the server with more information to work with than the first approach, enabling not only alerts to nearby users, but also to identify geographic hotspots and other patterns of contagion. It also provides a historic record of the evolution of an epidemic which can be mined and analyzed in many other ways.

Despite the simplicity of the second approach, privacy concerns have prevented its widespread adoption. Many people do not want their location history to be known to any third party, thus would avoid using any software that explicitly discloses this information. Some have gone so far as to call digital contact tracing based on unprotected disclosure of location data illegal or unconstitutional~\cite{TechRepublic:2020:DRA}. A number of commercial digital contact tracing apps, which report and store explicit location data, have been found in violation of user privacy policies, having shared this data with unauthorized third parties~\cite{MSN:2020:RFN}. Such privacy concerns must be addressed if digital contact tracing is to be deployed, as it is not very effective unless adopted by a majority of the population.

Technology that would enable the ``hiding'' or ``obfuscation'' of location data could go a long way towards alleviating privacy concerns and enabling contact tracing at a very large scale. Since the outbreak of \mbox{COVID-19}, this has been the topic of recent research, incorporating cryptographic techniques such as \emph{private set intersection}~\cite{Berke:2020:ADE}, \emph{private proximity testing} based on an \emph{equality testing protocol}~\cite{Fitzsimons:2020:ANO} and \emph{homomorphic encryption}~\cite{Bell:2020:TTP}. We refer the interested reader to the comprehensive surveys by Reichert et al.~\cite{Reichert:2020:ASO} and Messai and Seba~\cite{Messai:2020:PCO} on the privacy aspects of existing contact tracing apps.

The objective of this article is to describe a very simple method to hide data, which can also be used to hide spatio-temporal data. By hiding, we mean that instead of disclosing a data value $x$, a user would disclose an ``encoded'' version of $x$, namely $E(x)$. For this to be useful, it should be easy for any user to compute $E(x)$ if given $x$, but be very difficult, from a computational point of view, to invert $E$, namely to recover $x$ when provided only with $E(x)$ (even for the user who encoded $x$). By ``difficult'' we mean it would require a prohibitive amount of storage or of computational resources, which would effectively deter any such attempt. Although quite distinct, as we will make clear later, these resemble in spirit \emph{one-way functions} or \emph{cryptographic hash functions}~\cite{Katz:2020:ITM} used in classical cryptography. In its simplest form, the function $E$ is deterministic and injective, as then it is easy to check if $x=y$ by simply checking if $E(x)=E(y)$. In the contact-tracing scenario, the data $x=(t,l)$ is a data value consisting of a concatenation of the time $t$ with the location $l$. Given the function $E$, a user with ID~$i$ would periodically transmit to a central server the pair~$(i,e)$, where $e=E(x)$ is the encoded version of~$x$. The server would store these pairs in a database indexed by the second component. Given a \emph{query} vector $e$ (of a detected infection), it should be easy to search this database to determine all pairs $(i',e')$ such that $e'=e$, namely identify which other users (having~ID $i'$) were also at location $l$ at time $t$ and alert them.

We depart from traditional cryptographic techniques by not requiring the use of encryption keys of any sort, neither private nor public keys. This means that even the user who computed $E(x)$ from~$x$ cannot recover $x$ from $E(x)$ unless she explicitly records the connection between the two or stores some additional information which might facilitate the recovery. As we will see later, the security of the system follows from the sheer enormity of the relevant domain of spatio-temporal data (the so-called ``plaintext'' space) to be coded, which makes attacks on the system computationally infeasible. While the basic embodiment of $E$ is deterministic, it is possible to add an extra layer of security by introducing a non-deterministic (probabilistic) element to $E$, namely $E(x)$ could assume more than one value for any given $x$. In this case we need to modify the database search to a \emph{matching} procedure: given a query $e$, instead of searching for other vectors $e'$ such that $e=e'$, we search for all other vectors $e'$ such that $\delta(e,e')\le\tau$, where $\delta$ is the Hamming distance function between two vectors, namely the number of coordinates in which they differ, and $\tau$ is some threshold. These~$e'$ are called \emph{matches} of $e$. An exact match is, of course, the special case where $\tau=0$. A judicious choice of the encoding function $E$ and the value $\tau$ will guarantee no false negatives (i.e.\ perfect \emph{recall}), namely, given a query $e$ corresponding to some data $x$, we will always find \emph{all} other matching values~$e'$ corresponding to the same $x$. It will also guarantee a negligible (ideally zero) number of false positives (also called \emph{false alarms}), namely, almost never report values~$e'$ corresponding to a different data value $y\neq x$. In the contact tracing scenario, perfect recall is necessary so that \emph{all} individuals who have potentially been exposed to infection \emph{are alerted}. A tiny number of false positives are tolerable as all this means is that a small number of individuals who have not actually been exposed will be informed that they have.

This article proposes encoding functions for spatio-temporal data. In a nutshell, it maps a 2D location $l$ and time $t$, combined and represented as a large integer in a discrete world, to an $n$-dimensional vector of integers $E(x)$, where $n$ is quite large, e.g.\ $100$. The range of the components of $E(x)$ can be much larger than $n$, e.g.\ $\{0,\dots,502\}$. The function $E$ is based on well-known number-theoretic techniques, the preferred one making use of polynomials over finite fields. First deployed in 1960 in Reed--Solomon error-correcting codes~\cite{Reed:1960:PCO} and its variants (the most important being the BCH code), the technique has also found use in other cryptographic methods, such as Shamir's secret sharing method~\cite{Shamir:1979:HTS} and even blockchain~\cite{Cheng:2019:PMB}. The most important property of $E$ is that it transforms a very large integer into a long vector of much smaller integers in an injective way, which can be thought of as an \emph{embedding} in a higher-dimensional space, and this transformation cannot be inverted \emph{unless a minimal number $m\le n$ of the vector coordinates (and their indices in the vector) are known}. We take advantage of this by \emph{sorting} the vector coordinates so that their correspondence to the coordinate indices is lost, making it difficult to apply the standard decoding methods. An attacker has no choice but to try all possible permutations of subsets of size $m$ of the $n$ coordinates, making it computationally infeasible, even for relatively small values of $n$ and $m$. Another important property is that, although there are simple algebraic relationships between the coordinates of the vector, to the naked eye, and even to a statistical test, they look like random integers. Thus, the distribution of the encoded vectors in the embedding space is quite uniform, which will work in our favor.


\section{The setup}

Consider an integer domain $W=\{0,\dots,M-1\}$ (the ``world''). Any integer $x\in W$ is a valid (plaintext) message and we may express it as a sequence of $m$ digits $x=(x_1,\dots,x_m)$ in base $p$: $x=\sum_{i=0}^{m-1} x_i p^i$ where~$p$ is a prime number (or more generally a prime power) and $x_i\in\ZZ_p=\{0,\dots,p-1\}$. Note that this implies that~$m=\lceil\log_p M\rceil$ and taking a larger $m$ is superfluous. Essentially, $W$ is synonymous with a subset of $\ZZ_p^m$, the set of all vectors of length $m$, where each coordinate is taken from $\ZZ_p$.

In the contact tracing application, the spatio-temporal world consists of two-dimensional (latitude and longitude) GPS coordinates at 1 meter resolution (or the Open Location ``Plus Codes''~\cite{Google:2015:PCA}), which translates to a grid with $10^{14}$ points, and $10^5$ different time stamps for every $30$ seconds over the past month, implying a ``world'' of size $M=10^{19}$. If we use the prime $p=503$, this would mean $m=8$.


\section{The encoding function}

We propose the following non-deterministic encoding scheme:

\begin{center}\medskip
\fbox{\begin{minipage}[t]{.95\linewidth}
    Let $W=\{0,\dots,M-1\}$ be an integer domain, $n$ a positive integer and $p$ a prime. Denote by $\ZZ_p^n$ the set of vectors with $n$ elements from $\ZZ_p$ and by $\Delta_p^n$ the set of vectors with $n$ elements from $\ZZ_p$ in non-decreasing order, also known as the \emph{ordered discrete simplex}. The encoding function $E\colon W\to\mathcal{E}\subset\Delta_p^n$ has parameters $(M,p,n,k)$, where $0\le k\le n\le p$ and $n\ge m=\lceil\log_p M\rceil$.\medskip

    To compute $E(x)$ for a domain element $x\in W$:
    \begin{enumerate}
        \item Express $x$ in base $p$: $x=\sum_{i=0}^{m-1} x_i p^i$.
        \item Compute the \emph{basic} encoding \[C(x)=(\pi(0),\pi(1),\dots,\pi(n-1))\in\mathcal{C}\subset\ZZ_p^n,\] where $\pi(\xi)=\sum_{i=0}^{m-1} x_i \xi^i \pmod p$ is a polynomial of degree $m-1$ over the finite field $\ZZ_p$.
        \item Sort the coordinates of $C(x)$ in non-decreasing order to $C'(x)\in\mathcal{E}$.
        \item Randomly modify $k$ arbitrary coordinates of $C'(x)$, while preserving the increasing order of the coordinates, resulting in $E(x)\in\mathcal{E}$.
    \end{enumerate}
\end{minipage}}\medskip
\end{center}

\noindent Note that as a result of step (4), $k>0$ implies that $E(x)$ is non-deterministic, namely may assume multiple values.

The basic code \emph{space} $\mathcal{C}\subset\ZZ_p^n$, defined as the set of all possible basic codes of world elements $\mathcal{C}=\{C(x):x\in W\}$ consists of vectors of length $n$, such that $C_i(x)\in\ZZ_p$. It has the following properties:
\begin{enumerate}
    \item $C(x)$ is injective, namely $x=y$ if and only if $C(x)=C(y)$.
    \item $\mathcal{C}$ has \emph{Hamming distance} $d=n-m+1$, namely any two distinct codewords $c_1,c_2\in\mathcal{C}$ differ from each other by at least $d$ coordinates: $\delta(c_1,c_2)\ge d$. This is because any polynomial of degree $m-1$ over a field is uniquely determined by $m$ of its values. So not only is $C$ an injective function (i.e.\ $d>0$), but it maps distinct world elements quite far apart from each other in $\mathcal{C}$.
	\item $x$ may be recovered from $C(x)$ by a variety of efficient methods, including inverting a linear Vandermonde system~\cite[Section~6.1]{Horn:1991:TIM}.
\end{enumerate}
The basic coding function $C$ described above was proposed by Reed and Solomon~\cite{Reed:1960:PCO} as an error-correcting code to overcome corruption of $k=\lfloor d/2\rfloor$ coordinates of $C(x)$. When presented with $c'$, which is a corrupted version of $C(x)$, Property~2 guarantees that $C(x)$ is the unique codeword in $\mathcal{C}$ such that $\delta(C(x),c')\le k$, thus error-correction performed by replacing $c'$ with the vector closest to it in $\mathcal{C}$ by the Hamming distance, is well-defined and yields the correct result $C(x)$. The corrected codeword $C(x)$ may be found by efficient algorithms (e.g.~\cite{Gao:2003:ANA}), which take into account the special algebraic structure of $\mathcal{C}$.

Our non-deterministic encoding function is a variation on the theme of error-correction. In our scenario, we are presented with \emph{two} vectors $E(x), E(y)\in\mathcal{E}$ originating from $x,y\in W$. We would like to have a threshold~$\tau$ such that $x=y$ if and only if $\delta(E(x),E(y))\le\tau$.

To give the flavor of our approach, we remark that it is relatively easy to determine this threshold if the encoding procedure does \emph{not} contain the sorting step 3 in the encoding procedure, as the following lemma implies.

\begin{lemma}\label{lemma:threshold}
If we eliminate the sorting step 3 in the encoding procedure with parameters $(M,p,n,k)$, and set $k=\lfloor\frac{n-m}4\rfloor$ and $\tau=2k$ for $m=\lceil\log_p M\rceil$, then $x=y$ if and only if $\delta(E(x),E(y))\le\tau$.
\end{lemma}

\begin{proof}
From the definition of $k$, we have $n-m\ge4k$, so
\begin{align*}
    x = y
    &\quad\Rightarrow\quad C(x) = C(y)
     \quad\Rightarrow\quad \delta(C(x),C(y)) = 0\\
    &\quad\Rightarrow\quad \delta(E(x),E(y)) \le 2k = \tau,\\
    x \ne y
    &\quad\Rightarrow\quad \delta(C(x),C(y)) \ge n-m+1\\
    &\quad\Rightarrow\quad \delta(E(x),E(y)) \ge n-m+1-2k > 2k = \tau.
\end{align*}\vskip-1.5em
\end{proof}

\noindent
While not incorporating the sorting step 3 is amenable to easy analysis and identification of $k$ and $\tau$, it also compromises the security of the encoding $E(x)$, namely, it is then quite easy to recover $x$ from $E(x)$. This is essentially error-correction from $k$ errors, which, as mentioned above, is possible by a number of efficient algorithms, taking advantage of the special algebraic structure of $\mathcal{C}$~\cite{Gao:2003:ANA}.

The advantage of introducing sorting step 3 is precisely because it prevents the use of the standard error-correction algorithms, since the critical correspondence between the coordinates of $C'(x)$ (and thus of~$E(x)$) and the indices in the original $C(x)$ is lost.

The disadvantage of introducing sorting step 3 is that it modifies the Hamming distance $d$ present in $\mathcal{C}$, which is not likely to be preserved in $\mathcal{C'}$ and $\mathcal{E}$. In theory it could increase the distance, but it is much more likely to decrease it. It seems like it will be difficult to obtain a lower bound on this distance (which could have then been used to determine a threshold $\tau$, akin to Lemma~\ref{lemma:threshold}), since all the algebraic structure that was present in $\mathcal{C}$ has been destroyed in the transition to $\mathcal{C'}$ and $\mathcal{E}$.

Luckily, we are still able to make useful observations about the nature of the encoded vectors in $\mathcal{E}$. To the naked eye, the basic code space $\mathcal{C}$ will consist of integer vectors of essentially random values in the range~$\{0,\dots,p-1\}$. By ``random'' we mean actually pseudo-random, namely that although completely determined by $x$, it will be statistically impossible to distinguish between these vectors and completely random vectors. The sorting of the vectors will make them less random, but it will still be quite difficult to distinguish between the vectors in $\mathcal{E}$ and random non-decreasing integer vectors.


\section{The matching algorithm}\label{sec:matching}

Let us recall the application: We have a database of $D$ pairs of user ID's and encoded spatio-temporal values: $\{(i,E(x)):i=1,\dots,D\}$. Given the \emph{query} -- a vector $e$ -- we wish to find all \emph{matches} of $e$, namely, find all database entries $\{(i,e')\}$ such that both $e$ and $e'$ are possible encodings of the same data value $x$, i.e.\ $\delta(e,e')\le\tau$ for a suitable threshold $\tau$. We say that $\tau$ is the \emph{matching threshold} and $e'$ \emph{matches} $e$.

Recall that the size of the world is $M=\lvert W\rvert=10^{19}$. Assuming one billion (i.e.\ $10^9$) users, each storing location data for every $30$ seconds over the past month, namely, close to $10^5$ time-stamped locations, this implies that the database could contain $D=10^{14}$ entries.

We would like to show that even though the vectors are sorted, a matching threshold of $\tau=2k$ for ``reasonable'' values of $k$, as in Lemma~\ref{lemma:threshold}, is still a good choice. This is because the size of the database ($D$) is much smaller than the size of the world ($M$), thus the probability that database vectors match a typical query vector is infinitesimally small, unless they are encodings of the same world data.

Remember that $D\ll M\ll N$, where $M\approx p^m$ and $N=p^n$. Now, if given a query $e=E(x)$ for which there exists a matching database entry $e'$, then obviously $\delta(e,e')\le2k$. So to avoid false negatives, namely, to avoid missing correct matches, we must take $\tau\ge2k$.

Can we expect a given query vector $e=E(x)$ to ``accidentally'' match a vector $e'=E(y)$ corresponding to another $y\ne x$ in the database because of the sorting and corruption of the original basic code vectors in $\mathcal{C}$? The following theorem implies that this false positive is highly unlikely.

\begin{theorem}\label{theorem:probability}
Given any $e\in\mathcal{E}$, an upper bound for the probability of a vector $e'\in\mathcal{E}$, generated by sorting the coordinates of a random vector $z\in\ZZ_p^n$, differing from $e$ in at most $\tau$ non-adjacent coordinates is
\[
    \mathrm{Prob} \{ \delta(e,e') \le \tau \}
    \le s(p,n,\tau) = \frac{n!}{p^n} \sum_{d=0}^\tau \frac{{(2p)}^d}{d!}.
\]
\end{theorem}

\begin{proof}
For the case $\tau=0$, the probability of an exact match in all coordinates is at most $n!/p^n$, since all $n!$ permutations of $e$ can be taken as $z$ among all $p^n$ possible unsorted vectors in $\ZZ_p^n$, such that $\delta(e,e')=0$. For every coordinate of $e$ that occurs with multiplicity $\mu>1$, the probability reduces by a factor of $\mu!$, because the order of the repeated coordinate in $z$ does not matter.

For the case $\tau=1$, let us study the number of sorted vectors $e'\in\mathcal{E}$ that differ from $e$ in exactly one coordinate. Letting $e_0=0$ and $e_{n+1}=p-1$, it is clear that each coordinate $e_i'$ of $e'$ for $i=1,\dots,n$ can take any value in $\{e_{i-1},\dots,e_i-1,e_i+1,\dots,e_{i+1}\}$ without compromising the correct order. Hence, there are
\[
    \sum_{i=1}^n (e_{i+1}-e_{i-1}) = p - 1 + e_n - e_1 \le 2p - 2 \le 2p
\]
sorted vectors $e'\in\mathcal{E}$ at distance $\delta(e,e')=1$ from $e$ and thus the number of sorted vectors $e'\in\mathcal{E}$ with $\delta(e,e')\le1$ is at most $2p+1$. Using the same permutation argument as before, this proves the upper bound for $\tau=1$.

For the case $\tau>1$ we apply the previous argument iteratively $\tau$ times while using the assumption that the coordinates of $e'$ that differ from those of $e$ are non-adjacent. Then a vector at distance $\tau+1$ is just a modification of a vector at distance $\tau$ in one additional coordinate, thus the number of modifications is at most ${(2p)}^\tau$. Note that this is an overestimate as a modification may occasionally \emph{reduce} the distance by one. Since the order of modification of the modified coordinates is not important, we have counted each distinct modification $\tau!$ times.
\end{proof}

\noindent
The assumption that the differing coordinates of $e$ and $e'$ are non-adjacent makes the proof of Theorem~\ref{theorem:probability} easier, but we have experimentally observed that this upper bound holds also for the unrestricted case.

So the expected number of false positives for any given query $e$ is at most $D s(p,n,\tau)$, which decreases as $\tau$ decreases. For the values $p=503$, $n=100$, we may use $k=10$ and matching threshold $\tau=20$, thus $s(p,n,\tau)\approx10^{-71}$. Since $D=10^{14}$, the expected number of false positives per query is infinitesimal ($10^{-57}$), and even the expected number of false positives when each database entry is used as a query is still only $D^2 s(p,n,\tau)=10^{-43}$.

\subsection{Conclusion}
In our encoding scheme, it suffices to take a corruption parameter $k$ which is not too small and not too large, and then use $\tau=2k$ as the matching threshold. Such a threshold will completely avoid false negatives and produce a negligible number of false positives.

\subsection{Retrieving matching data}
Now that we have a suitable matching threshold for our matching algorithm, we must address the algorithmic question of how to organize the database of $D$ encoded values (which are sorted integer vectors), such that given a query vector~$e$, it is possible to efficiently find all pairs $(i',e')$ in the database such that $e'$ matches~$e$, namely such that $\delta(e,e')\le\tau$? This is known as the ``static Hamming distance range query''. Of course, exhaustive search of the database is possible, but that would cost $O(D)$ time, which is too costly in our scenario where $D=10^{14}$. Efficient data structures have been devised for dealing with this problem, as in Manku et al.~\cite{Manku:2007:DNF}. This requires $O(\tau nD)$ storage (which is significant but not prohibitive in our application) but has very fast ($O(\log D)$) query runtime. See also Liu et al.~\cite{Liu:2011:LSH} for more recent work on this problem.


\section{The tracing algorithm}  \label{sec:tracing_alg}

Now that we have an encoding algorithm and are able to match two encoded vectors, we describe the procedure to be followed by the individual users and the central server to do the actual contact tracing and alerts.

\subsection*{User with ID \boldmath$i$\unboldmath}
\begin{itemize}
    \item The user continuously transmits to the server data pairs $(i,e)$ where $e=E(x)$ and $x=(t,l)$ is her time and location, tagged as ``uninfected''. The user also stores the triples $(t,l,e)$ in a local database indexed by $t$ and $e$ (e.g.\ on her smartphone), so that it is easy to retrieve all $e$'s transmitted during a given time interval and recover $(t,l)$ from its encoding $e$.
    \item If the user discovers she is infected, she sends \emph{again} all pairs $(i,e)$  generated by her over the past, say, two weeks (by querying her local database) back to the server, tagged as ``infected''.
    \item Upon receipt of message $e$ tagged with ``possible infection'' from the server, the user recovers the infection time and location $(t,l)$ from $e$ (by querying her local database). The user self-isolates for two weeks and can possibly report $(t,l)$ separately to friends and family.
\end{itemize}

\subsection*{Central server}
\begin{itemize}
    \item Upon receipt of a data pair $(i,e)$ tagged ``uninfected'', the server stores the pair on the server database (of size $D$).
    \item Upon receipt of a pair $(i,e)$ tagged ``infected'', the server retrieves from the server database (by the matching algorithm described in Section~\ref{sec:matching}) all pairs $(i',e')$ for which $e'$ matches $e$. The server then sends these $e'$ to user $i'$ tagged with ``possible infection''.
\end{itemize}


\section{Attacking the code}\label{sec:attack}

Recall that a critical objective is to ``hide'' the data by its encoding, namely render it computationally infeasible to recover the (large) integer $y\in W$ from the integer vector $e=E(y)\in\mathcal{E}$, either because it would require too much computation time or too much storage space. We describe here three possible methods of attack and argue that they are infeasible. In all possible attacks, the large size $M$ of the spatio-temporal (plaintext) data domain  is what makes the attacks computationally impossible. Thus there is no need for an encryption key to further improve the system security.

\subsection{Brute-force attack}
The simplest method is just to exhaustively scan the entire world and check if the encoded version $e'=E(x)$ of any world point $x$ matches the given encoding $e$ (namely, that $\delta(e,e')\le\tau$). This would require $\lvert W\rvert=M=10^{19}$ encodings and comparisons, which is prohibitive in runtime.

\subsection{Table attack}
We could reduce the runtime of the brute-force attack by trading off space for time, employing a very large database. Simply compute some encoding $E(x)$ for every possible $x\in W$ in a preprocessing phase and store the pairs $(x,E(x))$ in a database indexed by $E(x)$. Given an encoding $e$, the matching algorithm described in Section~\ref{sec:matching} would then be able to quickly retrieve all matches of $e$. However, this requires a database of size $O(\tau nM)$ which is $\frac{M}{D}=10^5$ times larger than the server database. For $M=10^{19}$, $p=503$, $n=100$, and $\tau=20$, this is at least $10^{21}$ bytes, and would be prohibitively large.

\subsection{Direct attack}\label{sec:direct_attack}
A direct attack occurs when an adversary tries to invert the encoding through a subset of the coordinates by applying the traditional decoding algorithms such as solving a linear Vandermonde system. This is foiled by the sorting of the coordinates of the vectors. Since inversion requires knowledge of the correspondence between coordinates and their indices for at least $m$ uncorrupted coordinates, this is what an attempt to invert $e=E(x)$ must look like:

\begin{center}\medskip
\fbox{\begin{minipage}[t]{.95\linewidth}
	\begin{algorithmic}[1]
    \For{each of the $\binom{n}{m}$ subsets of $m$ coordinates}
      \For{each of the $\frac{n!}{(n-m)!}$ permutations of $m$ indices}
        \State solve for $x$ \Comment{\parbox[t]{0.75\linewidth}{\small e.g., by multiplying $e$ by the inverse of the Vandermonde sub-matrix consisting of the corresponding $m$ rows from the full $n\times m$ Vandermonde matrix}}
        \If{$x\ge M$}
          \State continue
        \EndIf
        \State compute $e'=E(x)$
        \If{$\delta(e,e')\le\tau$}
          \State return($x$)
        \EndIf
      \EndFor
    \EndFor
	\end{algorithmic}
\end{minipage}}\medskip
\end{center}

\noindent Each solve costs $\Omega(m)$ time. Should any of the selected subset of $m$ coordinates be corrupted, the inner loop will run completely, costing $\frac{n!}{(n-m)!}$ solves. Since the probability that none of the $m$ coordinates are corrupted is $\bigl(1-\frac{k}{n}\bigr)^m\approx\exp\bigl(-\frac{km}{n}\bigr)$, the outer loop will terminate on the average after $\exp\bigl(\frac{km}{n}\bigr)$ iterations and the inner loop will compute an expected number of $\frac{n!}{2(n-m)!}$ solves the last time it runs. Note that failure in one iteration due to one or more corrupted coordinates will not reveal which of the $m$ coordinates are corrupted, so that there is no extra information that can help to choose a ``better'' set of $m$ coordinates in the next iteration. In total, the expected number of solves for this attack would be $\frac{n!}{(n-m)!}\exp\bigl(\frac{km}{n}\bigr)$. For $n=100$ and $p=503$, we have $m=8$. With $k=10$, the expected number of solves is $10^{16}$, which would take too long.


\section{An alternative: redundant residue number systems}

While we have presented an encoding method based on polynomials over finite fields, it is possible to use another method which is also employed in error-correcting coding and secret-sharing. This involves so-called \emph{redundant residue number systems}. Originally proposed in the 1950's for efficient arithmetic computations on large integers~\cite{Garner:1959:TRS}, this technique was adopted for error-correction coding soon after~\cite{Watson:1966:SCU,Barsi:1973:ECP} and is also used in cryptography~\cite{Mignotte:1983:HTS,Asmuth:1983:AMA}. The main difference between this method and the basic coding method described above based on polynomials is that now the basic code space is $\mathcal{C}=\ZZ_{p_1}\times\ZZ_{p_2}\times\dots\times\ZZ_{p_n}$ for a sequence of distinct primes $(p_1,\dots,p_n)$, instead of $\ZZ_p^n$.

\begin{table}\centering\small
  \newlength{\colA}\settowidth{\colA}{polynomial}
  \newlength{\colB}\settowidth{\colB}{encoded}
  \newlength{\colC}\settowidth{\colC}{alphabet}
  \newlength{\colD}\settowidth{\colD}{$m=\lceil\log_p M\rceil$}
  \newlength{\colE}\settowidth{\colE}{coordinates}
  \newlength{\colF}\settowidth{\colF}{threshold}
  \newlength{\colG}\settowidth{\colG}{size (bits)} 
  \newlength{\colH}\settowidth{\colH}{$D^2 s(n,p,\tau)$}
  \newlength{\colI}\settowidth{\colI}{$\frac{n!}{(n-m)!}\exp\bigl(\frac{km}{n}\bigr)$}
  \tabcolsep2.3pt
  \begin{tabular}{
        p{\colA}
        >{\centering\let\newline\\\arraybackslash\hspace{0pt}}p{\colB}
        >{\centering\let\newline\\\arraybackslash\hspace{0pt}}p{\colC}
        >{\centering\let\newline\\\arraybackslash\hspace{0pt}}p{\colD}
        >{\centering\let\newline\\\arraybackslash\hspace{0pt}}p{\colE}
        >{\centering\let\newline\\\arraybackslash\hspace{0pt}}p{\colF}
        >{\centering\let\newline\\\arraybackslash\hspace{0pt}}p{\colG}
        >{\centering\let\newline\\\arraybackslash\hspace{0pt}}p{\colH}
        >{\centering\let\newline\\\arraybackslash\hspace{0pt}}p{\colI}}
    \toprule
    coding method &
    encoded vector length &
    alphabet size &
    data vector size (base $p$) &
    corrupted coordinates &
    matching threshold &
    encoded vector size (bits) &
    expected\par \# of false positives &
    ``direct attack'' complexity\\
    & $n$\rule{0pt}{3ex} & $p$ & $m=\lceil\log_p M\rceil$ & $k$ & $\tau$ &
    $\lceil n\log_2 p\rceil$ & $D^2 s(n,p,\tau)$ &
    $\frac{n!}{(n-m)!}\exp\bigl(\frac{km}{n}\bigr)$\\
    \midrule
    polynomial & $100$ & $503$ & $8$ & $10$ & $20$ & $898$ & $10^{-43}$ & $10^{16}$\\
    polynomial & $100$ & $101$ & $10$ & $1$ & $2$ & $666$ & $10^{-10}$ & $10^{20}$\\
    polynomial & $200$ & $211$ & $9$ & $20$ & $40$ & $1545$ & $10^{-5\phantom{0}}$ & $10^{21}$\\
    residues & $80$ & $\approx1143$ & $7$ & $8$ & $16$ & $858$ & $10^{-58}$ & $10^{13}$\\
    \bottomrule
  \end{tabular}
  \caption{Parameters of the different settings for world size $M=10^{19}$ and database entries $D=10^{14}$.}
  \label{tab:examples}
\end{table}

Recall that the ``world'' is $W=\{0,\dots,M-1\}$. Let $(p_1,\dots,p_n)$ be a sequence of increasing primes, $m$ an integer such that $\prod_{i=n-m+2}^n p_i<M<\prod_{i=1}^m p_i$, and denote $N=\prod_{i=1}^n p_i$. The encoding function $E\colon W\to\mathcal{E}$ for a domain element $x\in W$, has parameters $(p_1,\dots,p_n,k,n)$, where $p_i$ are primes and $0\le k\le n$ is an integer. The basic coding function is simply $C(x)=(x \pmod {p_1},\dots,x \pmod {p_n})\in\mathcal{C}$. Similar to the case of polynomials over finite fields, the infamous Chinese Remainder Theorem~\cite{Barsi:1973:ECP} guarantees that $x$ can be recovered from any subset of $m$ coordinates of $C(x)$ along with their indices, so this code also has Hamming distance $n-m+1$, and error-correction may be done using a variety of methods taking advantage of the algebraic structure (e.g.~\cite{Goldreich:2000:CRW}). Our encoding proceeds as above, by sorting the coordinates of the basic code and corrupting a small subset without changing the order. Nothing else is changed.

Despite this approach actually being simpler to implement than the polynomial-based approach, it is less desirable due to $m$ being more constrained as a function of the primes used. For example, for $n=80$, taking as $p_i$ all the consecutive primes from $877$ to $1,451$ (having geometric mean $1,143$) yields only $m=7$. An appropriate $k$ would be $8$, thus $\tau=16$. The probability of a false positive is then $10^{-58}$ and the complexity of the direct attack is $10^{13}$ (see Table~\ref{tab:examples}).


\section{Discussion and extensions}

We have described just a very basic version of a possible contact-tracing system, where our main contribution is the data coding method, which  plays a central and critical role
(and could be useful for other applications). A more realistic system may require more than just this simple feature set. In this section we describe a number of possible extensions that could make our system more applicable to a real-world setting.

\subsection{Increasing the security}
It is relatively easy to increase the security of the system, i.e.\ making a direct attack on the system more difficult. In the scenario described above, where $M=10^{19}$, we took $n=100$, $p=503$, implying $m=8$, thus the complexity of a direct attack is $10^{16}$. If we were to take instead $n=100$ and $p=101$, so that $m=\bigl\lceil\log_{101}10^{19}\bigr\rceil=10$, the complexity would increase to $10^{20}$ (although we would have to take $k=1$ and $\tau=2$ to keep the probability of a false positive at $10^{-10}$), and if this were not enough, we can increase this further by increasing both $n$ and $p$. See Table~\ref{tab:examples} for a comparison of the attack complexity resulting from different values of the system parameters. Increasing $n$ obviously increases the (bit) size $n\log_2 p$ of the code $C(x)$ and thus the size of the server database, but the same is true for the database of the ``table attack''.

\subsection{Using a deterministic mapping}
Our encoding method is non-deterministic, namely involves randomly corrupting a subset of $k>0$ coordinates in the sorted basic code vector. The advantage of a large $k$ is that it increases the difficulty of a direct attack on the database, as described in Section~\ref{sec:direct_attack}. However, for certain values of the other system parameters, it may be possible to make do with a deterministic encoding method, namely $k=\tau=0$. In this case, matching a query vector within the server database reduces to exact vector match, which may be done easily by binary search on a table (of size $D$) of the database entries $(i,e)$, sorted in lexicographic order of $e$.

\subsection{Detecting persistence in time}
A common assumption for potential infection is temporal \emph{persistence}, i.e.\ continuous exposure for a significant amount of time (typically 15 minutes). The basic embodiment of our system detects and alerts for contact at a specific point in time (and space), however it is straightforward to extend it to deal with persistence. This is done client-side, namely by the user. Exposure to infection for $k$ consecutive  time stamps will result in $k$ alerts to the user, at which point she can check for herself (in the third bullet of the user algorithm in Section \ref{sec:tracing_alg}) for the temporal persistence of these alerts and proceed accordingly.

\subsection{Detecting proximity in space}
The method outlined in this article provides an easy way to determine whether $x=y$ by comparing $E(x)$ and $E(y)$. Recall that $x$ and $y$ are taken from a discrete world, which are essentially samples of the true continuous world at some finite resolution grid. However, sometimes in contact tracing it is necessary to also determine proximity beyond the grid resolution, either because of an increased radius of infection or simply because the accuracy of the measured location (typically taken from a GPS device) is much worse than the grid resolution and the chances of an exact match in measured location even when two users are within grid resolution, is very slim.

It would seem difficult to achieve this, since the encoded vectors have a pseudo-random distribution and any spatio-temporal correlation between two data points would be ``lost in encoding''. The easy way to circumvent this is for the user to transmit to the central server encodings of not just her current location, but also of the neighboring grid points, effectively ``dilating'' the data point. This would incur some modest overhead in storage and transmission costs on both client-side and server-side.

\subsection{Server-side analytics}
Reporting absolute locations has the advantage that the server can run analytics on the accumulated data, e.g.\ to detect spatio-temporal infection ``hotspots'' or other contagion patterns over time and space. However, this requires the server to access the unencoded (time, location) data vectors reported by the users after infection, a feature that our basic system does not support. One way around this, while maintaining user anonymity, is that the user, upon detecting infection, additionally reports to the server the unencoded data $x$ (\emph{without} the user ID $i$) using a separate protocol that guarantees anonymity of the sender. This would be added to the second bullet of the user algorithm in Section \ref{sec:tracing_alg}.

\subsection{``Inflating'' the world}
The world size, in our contact tracing application, is $M=10^{19}$ integers, which is very large, but constrains some of the parameters in our encoding scheme. In particular, the parameter $m$, if too small, could compromise the security against the direct attack, as described in Section~\ref{sec:direct_attack}. One way to rectify this would be to ``inflate'' the world by means of some function $f\colon W\to W'$ with $M=\lvert W\rvert\ll\lvert W'\rvert=M'$. This function $f$ should be injective and non-polynomial, so that it cannot be inverted easily at each individual coordinate. One possibility for such an $f$ is the following:

Let $q_i$ denote the $i$-th prime (i.e.\ $q_1=2$, $q_2=3$, etc.) and observe that the product of the first $m_0=16$ primes is a little larger than the size of our world. Hence, the first step is to map $x\in W$ to the residue code vector w.r.t.\ these $16$ primes, namely compute $C(x)=(c_1,\dots,c_{m_0})$ with $c_i=x \pmod {q_i}$. For the next step, let $s_i=\sum_{j=1}^{i-1} q_i$ denote the sum of the first $i-1$ primes (i.e.\ $s_1=0$, $s_2=2$, $s_3=5$, etc.) and let us map each $c_i$ to the $(s_i+c_i+1)$-th prime, giving the vector $C'(x)=(c_1',\dots,c_{m_0}')$ with $c_i'=q_{s_i+c_i+1}$. Finally, we define $f(x)=\prod_{i=1}^{m_0}c_i'$ and note that $f(x)$ is a \emph{square-free} integer with exactly $m_0$ prime factors. Moreover, as the mapping $C$ is injective, it follows that $f(x)$ and $f(y)$ for $x\ne y$ have at most $m_0-1$ common factors, thus guaranteeing the injectivity of $f$. The size of the inflated world is $M'=\prod_{i=1}^{m_0} q_{s_{i+1}}\approx10^{39}$. We now continue to encode $x'=f(x)\in W'$ instead of $x\in W$ with the polynomial-based approach outlined above, but now having the advantage of a larger $m'=15$ instead of the previous $m=8$.

\subsection{Other linear codes}
The basic code based on polynomials that we use is a linear code, in the sense that the coding operation is just multiplication by a matrix: $C(x)=V x$ over $\ZZ_p$. $V$ is the $n\times m$ Vandermonde matrix, which has the special property that \emph{all} submatrices of size $m\times m$ have full rank. This property allows to recover $x$ from \emph{any} subset of $m$ coordinates of $C(x)$ by multiplying them by the inverse of the appropriate submatrix of $V$. Thus \emph{any} $n\times m$ matrix with similar properties would serve the same purpose. Furthermore, were we to construct an $n\times m$ matrix $A$ with the property that \emph{some} of the submatrices of size $m\times m$ have rank less than $m$, and that full rank is obtainable only when the submatrix is enlarged to $(m+l)\times m$, this, coupled with the corruption of coordinates during encoding, could further complicate the direct attack on the method described in Section~\ref{sec:direct_attack}.


\end{document}